\documentclass[aps,prb,twocolumn,superscriptaddress]{revtex4-2}

\usepackage{amssymb}
\usepackage{amsmath}
\usepackage{graphicx}

\usepackage[colorlinks,linkcolor=blue,anchorcolor=blue,citecolor=red]{hyperref} 

\usepackage{array}
\usepackage{multirow}
\usepackage{setspace}

\begin{document}

\title{DFT+DMFT investigation of the magnetic phase transition in the itinerant ferromagnet Fe$_{3}$GaTe$_{2}$}
\author{Yuanji Xu}
\email{yuanjixu@ustb.edu.cn}
\affiliation{Institute for Applied Physics, University of Science and Technology Beijing, Beijing 100083, China}
\author{Xintao Jin}
\affiliation{Institute for Applied Physics, University of Science and Technology Beijing, Beijing 100083, China}
\author{Jiacheng Xiang}
\affiliation{Institute for Applied Physics, University of Science and Technology Beijing, Beijing 100083, China}
\author{Huiyuan Zhang}
\affiliation{Institute for Applied Physics, University of Science and Technology Beijing, Beijing 100083, China}
\author{Fuyang Tian}
\email{fuyang@ustb.edu.cn}
\affiliation{Institute for Applied Physics, University of Science and Technology Beijing, Beijing 100083, China}
\date{\today}

\begin{abstract}
Finding and designing ferromagnets that operate above room temperature is crucial in advancing high-performance spintronic devices. The pioneering van der Waals (vdW) ferromagnet Fe$_{3}$GaTe$_{2}$ has extended the way for spintronic applications by achieving a record-high Curie temperature among its analogues. However, the physical mechanism of increasing Curie temperature in this material still needs to be explored. Here, we systematically investigate the electronic structures and magnetic properties of Fe$_{3}$GaTe$_{2}$ as a function of temperature using strongly correlated calculations, reconciling the dual nature of $d$-electrons with both localization and itinerant characters. Significantly, our study reveals the emergence of quasi-particle flat bands driven by many-body interactions, which enhance magnetic stability through a positive feedback mechanism. Furthermore, our results demonstrate the hybridization of these flat bands at low temperatures, indicating the possible presence of heavy fermion behavior in this system. Our findings suggest that tunable flat bands near the Fermi level may serve as a key factor in realizing materials with high magnetic transition temperatures and strong magnetic anisotropy. This research provides a promising pathway for exploring next-generation spintronic devices utilizing vdW flat band systems.
\end{abstract}

\maketitle

\section{INTRODUCTION}

Two-dimensional (2D) quantum magnetic systems represent an excellent platform for spintronic applications \cite{Sarkar2015,Liu2020,Yang2022,Gong2019}. Particularly noteworthy is the recently discovered itinerant vdW ferromagnet Fe$_{3}$GaTe$_{2}$, which exhibits a remarkable ferromagnetic (FM) phase transition temperature exceeding 350 K, positioning vdW materials as promising candidates for advancing spintronic technologies \cite{Zhang2022}. Notably, Fe$_{3}$GaTe$_{2}$ not only boasts a significantly higher Curie temperature ($T_{\rm C}$) compared to its counterpart Fe$_{3}$GeTe$_{2}$ \cite{Fei2018,Deng2018}, but also demonstrates robust perpendicular magnetic anisotropy (PMA) \cite{Deng2018,Ikeda2010,Gong2017}. Despite these advancements, the underlying physical mechanisms driving the high-$T_{\rm C}$ behavior remain incompletely understood. Recent theoretical investigations have made strides in elucidating the origin of this phenomenon, estimating exchange interactions through density functional theory (DFT) calculations combined with a Heisenberg-type Hamiltonian \cite{Ghosh2023,Lee2023,Ruiz2024}. These studies reveal that the magnetic interactions in Fe$_{3}$GaTe$_{2}$ and Fe$_{3}$GeTe$_{2}$ are quite intricate, with in-plane couplings playing a pivotal role in differentiating their $T_{\rm C}$ \cite{Ruiz2024}.

\begin{figure}
\begin{center}
\includegraphics[width=0.49\textwidth]{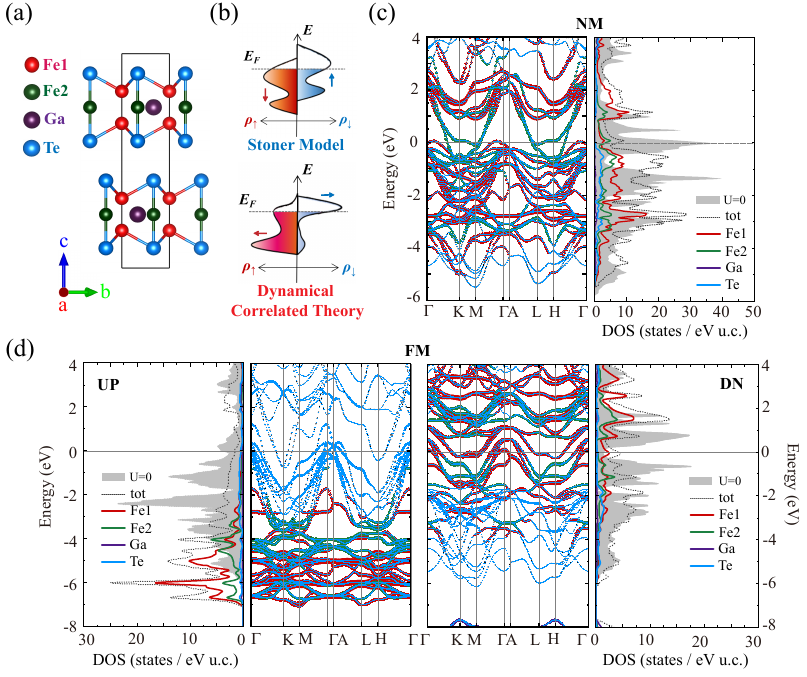}\caption{(a) Illustration of the crystal structure of Fe$_{3}$GaTe$_{2}$. (b) The recent proposed paradigm depicting changes in the density of states during magnetic transition in correlated itinerant ferromagnets \cite{Xu2024}. (c) Electronic structures and partial density of states from non-spin-polarized calculations of Fe$_{3}$GaTe$_{2}$ in DFT+U calculations. (d) Electronic structures and partial density of states of ferromagnetic order in DFT+U calculations. The gray areas in (c) and (d) mean the total density of states with $U=0$ eV.}
\label{fig1}
\end{center}
\end{figure}

However, addressing magnetism in itinerant ferromagnets has historically posed challenges due to the dual nature of $d$-electrons \cite{Yin2011,Georges2013,Georges2024,Bao2022}. Experimental and theoretical studies have underscored the critical role of strongly correlated effects in understanding magnetic transitions in 2D vdW ferromagnets \cite{Zhu2016,Zhang2018,Zhao2021,Corasaniti2020}. For instance, previous experiments have observed significant electron mass enhancement in Fe$_{3}$GeTe$_{2}$ \cite{Zhu2016,Zhang2018,Zhao2021,Chen2013}, which shares the same crystal structure as Fe$_{3}$GaTe$_{2}$ shown in Fig.~\ref{fig1}(a). Additionally, angle-resolved photoemission spectroscopy (ARPES) experiments have revealed weak temperature-dependent shifts in bands, indicating that neither the Heisenberg localized model nor the Stoner itinerant model alone can adequately describe itinerant ferromagnets \cite{Xu2020,Wu2024}. Recently, Xu and Tian proposed the dynamical correlated model, which accurately accounts for the magnetic phase transition through the spectral weight transfer process, as depicted in Fig.~\ref{fig1}(b) \cite{Xu2024}. Their theoretical investigation of Fe$_{3}$GeTe$_{2}$ has revealed the emergence of flat bands near the Fermi level. As flat bands are recognized to exert a significant influence on magnetism, heavy fermions, superconductivity and topology \cite{Rosa2024,Huang2024,Checkelsky2024}, this perspective strongly suggests a potential novel approach to exploring high-$T_{\rm C}$ candidate ferromagnetic spintronic materials based on flat band characteristics.

In this study, we simulate the magnetic phase transition of Fe$_{3}$GaTe$_{2}$ using density functional theory combined with dynamical mean-field theory (DFT+DMFT) \cite{Georges1996,Kotliar2006}. Our calculations yield a magnetic transition above room temperature and avoid the controversy about using the local Heisenberg model to estimate the Curie temperature in itinerant systems. Additionally, our calculations exhibit Curie-Weiss behavior of spin susceptibility, which can not be captured by the Stoner model \cite{Stoner1947}. We observe spectral weight transfer between the two spin channels in Fe$_{3}$GaTe$_{2}$, highlighting the crucial role of correlated effects. Our findings demonstrate the emergence of flat bands at low temperatures, where the development of long-range magnetic order effectively suppresses spin fluctuations and facilitates flat band formation. Moreover, our calculations suggest that these flat bands exhibit heavy fermion characteristics. Besides, the flat bands near the Fermi level can be finely tuned, providing a greater opportunity for the realization of materials with high magnetic transition temperatures and strong magnetic anisotropy. Our study suggests that flat band systems have significant potential as platforms for practical spintronic devices.

\section{METHODS}

First-principles calculations were performed using the full-potential linearized augmented plane-wave (FLAPW) method, as implemented in the WIEN2k package \cite{Blaha2023}. The Perdew–Burke–Ernzerhof (PBE) functional was employed to represent the exchange-correlation potential \cite{Perdew1996}. In our calculations, we focused on simulating the bulk material, as previous studies have indicated that the underlying physics remains similar between bulk and single-layer structures \cite{Zhang2022}. Additionally, we used the VASP code to relax the lattice parameters and ionic positions in Fe$_{3}$GaTe$_{2}$, employing the projector augmented-wave method \cite{Kresse1996,Blochl1994}. The van der Waals interactions were included using the DFT-D3 method of Grimme \cite{Grimme2010}. The optimized structure yields lattice parameters of $a=4.034\,$\AA{}, $c=16.0896\,$\AA{}, with the Fe1 position at (0, 0, 0.67519) and the Te position at (0.6667, 0.3333, 0.59269). The muffin-tin radii were set to 2.26 a.u. for Fe, 2.06 a.u. for Ga; and 2.23 a.u. for Te. The k-point grid was set $16 \times 16 \times 3$ for the bulk, and the product of the atomic sphere radius and the plane-wave cutoff R$_{\rm mt}$K$_{\rm max}$ was chosen to be 8. To account for the strong correlation effects in transition metal elements, an effective Coulomb repulsion $U_{\rm eff}$ of 5.0 eV was applied to the Fe atoms within the GGA+U framework, utilizing the self-interaction correction method introduced by Anisimov \cite{Anisimov1993}.

To account for the dynamical electronic correlations in Fe-$d$ orbitals, we performed fully self-consistent DFT+DMFT calculations \cite{Georges1996,Kotliar2006}. These calculations were carried out using the EDMFT code developed by Haule et al. \cite{Haule2010}. A large energy window, spanning from -10 eV to +10 eV relative to the Fermi level, was used to construct the DMFT projectors. The double-counting correction for the self-energy was applied using the nominal scheme. The hybridization expansion continuous-time quantum Monte Carlo (CT-HYB) method was employed as the impurity solver \cite{Werner2006,Haule2007}. The number of Monte Carlo sweeps is about $3 \times 10^{8}$. We adopted an Ising-type Hund’s coupling, with correlation parameters set to $U=$ 5.5 eV and $J=$ 0.7 eV. These values are consistent with previous studies on its sister compound \cite{Zhu2016,Xu2024} and ensure that the converged Fe atomic occupation remains in close agreement with experimental measurements \cite{Zha2023}. For the magnetic DFT+DMFT calculations, we considered all 10 Fe-$d$ orbitals per Fe atoms. To initialize the ferromagnetic state, a stepwise energy difference was introduced between the impurity levels of the spin-up and spin-down channels in the initial self-energy. For simplicity, we fixed the volume across different temperatures. Since the optimized lattice parameters obtained from DFT calculations vary by less than 2\% due to magnetism, and the magnetic moment exhibits only minor changes with these small variations in lattice parameters. This approximation is expected to have a negligible impact on the main physical conclusions of our study.

\section{RESULTS AND DISCUSSION}

\begin{figure}[b]
\begin{center}
\includegraphics[width=0.40\textwidth]{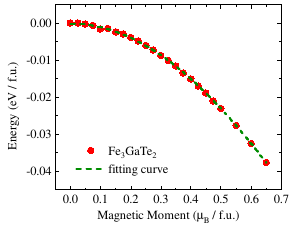}
\caption{Fixed moment DFT calculation of the total energy as a function of constrained spin magnetization for the Fe$_{3}$GaTe$_{2}$. The energies are given relative to the non-magnetic calculation.}
\label{fig2}
\end{center}
\end{figure}

\begin{table*}
\caption{\label{tab1}Magnetic moments obtained from DFT, DFT+U and DFT+DMFT (50 K) calculations, along with local magnetic moments in both ferromagnetic (50 K) and paramagnetic (1000 K) DFT+DMFT calculations. In DFT+DMFT calculations, the local magnetic moment is computed as the sum of the probabilities multiplied by the absolute value of the spin magnetic moment for each state from the impurity solver.\\}
\centering
\begin{spacing}{1.4}
\begin{tabular}{p{1.8cm}<{\centering}|p{2.5cm}<{\centering}|p{2.5cm}<{\centering}|p{2.5cm}<{\centering}|p{3.5cm}<{\centering}|p{3.5cm}<{\centering}}
\hline
\hline
\multirow{2}{*}{}  & \multicolumn{3}{|c}{Magnetic Moment} & \multicolumn{2}{|c}{Local Magnetic Moment}  \\
\cline{2-6}
                                                     &    DFT    &     DFT+U   &    DFT+DMFT   & DFT+DMFT (FM)  & DFT+DMFT (PM)  \\
\hline
$M_{\rm Fe1}$ ($\mu_{\rm B}$)   &    2.31    &     3.22        &         2.62           &             3.13          & 3.13  \\
\hline
$M_{\rm Fe2}$ ($\mu_{\rm B}$)  &    1.50     &     2.62        &          0.62         &               1.66         & 1.63  \\
\hline
\hline
\end{tabular}
\end{spacing}
\end{table*}

Fe$_{3}$GaTe$_{2}$ is a layered vdW ferromagnet with two inequivalent Fe sites, designated as Fe1 and Fe2, as depicted in Fig.~\ref{fig1}(a)  \cite{Zhang2022}. To investigate the origin of high-$T_{\rm C}$ in Fe$_{3}$GaTe$_{2}$, the Stoner model, commonly used in itinerant systems, is employed \cite{Chen2013}. To elucidate the electronic structure and estimate the Stoner parameter, we performed non-magnetic DFT calculations on the relaxed bulk crystal structure. As shown in Fig.~\ref{fig1}(c),  the non-magnetic (NM) DFT ($U=$ 0 eV) results (represented by the gray area) reveal a considerable DOS at the Fermi level, consistent with the Stoner model. To estimate the Stoner parameters $I$, we conducted fixed spin moment (FSM) calculations by fitting the toal energy $E$ as a function of the magnetic moment  $M$ using a polynomial expansion $E(M)=E_{0}+a \times M^{2}+b \times M^{4}$, where $a$ is related to the Stoner parameter $I$ via $a=1/N(E_{\rm F})-I$, with $N(E_{\rm F})$ denoting the non-magnetic calculated DOS at the Fermi level \cite{Kubler2009,Kakahashi2013,Sieberer2006,Qiao2023}. As shown in Fig.~\ref{fig2}, out fitting yields $a=$ -0.097 eV and $b=$ 0.018 eV in DFT calculations with $U=$ 0 eV. Using $N(E_{\rm F})=$ 16.15 states/ eV f.u., the Stoner parameter $I$ is determined to be 0.1586 eV. Consequently, Stoner's criterion of $I \times N(E_{\rm F})= 2.56 > 1$ is satisfied, indicating the material’s tendency towards ferromagnetism.

Additionally, previous experimental and theoretical studies have demonstrated that Fe$_{3}$GaTe$_{2}$ and its isostructural sister compound, Fe$_{3}$GeTe$_{2}$, exhibit strong correlation effects \cite{Zhu2016,Wu2024,Xu2020,Xu2024}. Base on these findings, we further conducted DFT+U calculations to account for strongly electronic correlations, incorporating an effective Coulomb repulsion of $U_{\rm eff}=$ 5.0 eV. While standard DFT calculations predict a large DOS at the Fermi level, our DFT+U calculations for the non-magnetic state reveal a significantly suppression of the DOS at the Fermi level, as shown in Fig.~\ref{fig1}(c). Moreover, the electronic structures in the ferromagnetic state obtained from both DFT and DFT+U calculations exhibit substantial band spin-splitting compared to the non-magnetic results, as illustrated in Fig.~\ref{fig1}(d).  As a result, neither the DFT nor DFT+U calculation can explain the weak temperature-dependent band shift observed in experiments \cite{Wu2024}. Given that the heavy fermion behavior and Kondo physics reported in its sister compound, Fe$_{3}$GeTe$_{2}$ \cite{Zhu2016,Zhang2018,Zhao2021,Xu2024}, these findings suggest that dynamical correlation effects should be considered in Fe$_{3}$GaTe$_{2}$ when investigating its magnetic properties.

\begin{figure*}
\begin{center}
\includegraphics[width=0.90\textwidth]{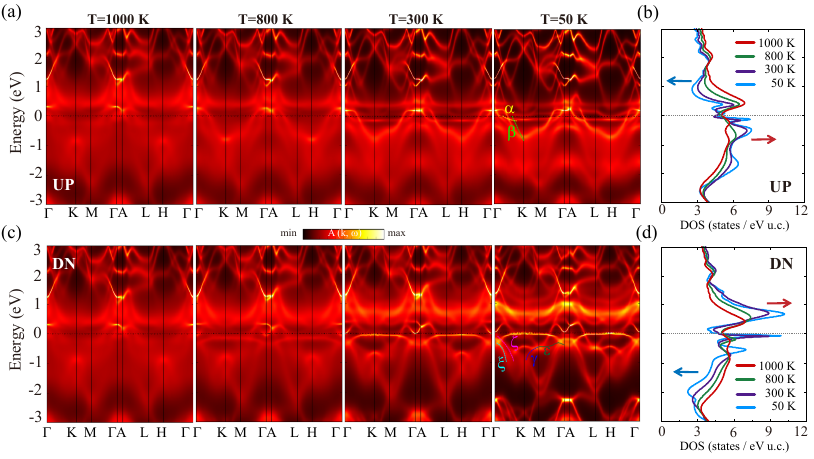}
\caption{The DFT+DMFT results of Fe$_{3}$GaTe$_{2}$ with varying temperatures. The calculated spectral functions of the spin-up (a) and spin-down (c) channels are shown for temperatures ranging from 1000 K to 50 K. The colored dashed lines represent the experimental band structures extracted from ARPES measurements \cite{Wu2024,Lee2023}. For clarity and ease of comparison, the band labels $\alpha$, $\beta$, $\gamma$, $\xi$, and $\zeta$ follow the notation used in Wu's work \cite{Wu2024}, and the band label $\varepsilon$ corresponds to the notation in Lee's work \cite{Lee2023}. The corresponding density of states of the spin-up (b) and spin-down (d) channels are also shown for temperatures ranging from 1000 K to 50 K. Blue arrows indicate areas where the density of states decreases with decreasing temperature, while red arrows indicate areas where the density of states increases with decreasing temperature.}
\label{fig3}
\end{center}
\end{figure*}

To simulate the magnetic phase transition directly, we further performed magnetic DFT+DMFT calculations, which have yielded reliable results consistent with experimental findings in its sister system, Fe$_{3}$GeTe$_{2}$ \cite{Xu2024,Kim2022,Bai2022}. Figure~\ref{fig3} illustrates the evolution of electronic structures as a function of temperature. At high temperatures, such as 1000 K, the electronic structures of the spin-up and spin-down channels are identical, and the system converges to the paramagnetic (PM) state. The overall DOS at high temperatures is broad, accompanied by large blurred regions in the spectral function. These incoherent characteristics originate from significant spin fluctuations of Fe-$d$ electrons. Notably, the system undergoes spontaneous symmetry breaking of magnetism at low temperatures, converging to the ferromagnetic order. As shown in Figs.~\ref{fig3}(a) and \ref{fig3}(c), the spectral functions of the spin-up and spin-down channels exhibit distinct differences as the temperature decreases. Correspondingly, the DOS in Figs.~\ref{fig3}(b) and \ref{fig3}(d) show the spectral weight transfer in different tendencies. The spectral weight above the Fermi level in the spin-up channel decreases as temperature decreases, while it increases below the Fermi level as temperature decreases. Conversely, the situation is the opposite in the spin-down channel.

In recent DFT+DMFT study of its sister material, Fe$_{3}$GeTe$_{2}$, the dynamical correlated effects have been shown to be crucial for accurately capturing the electronic structures \cite{Xu2024}. Evidently, the spectral weight transfer process and the little shift of bands near the Fermi level are also observed in Fe$_{3}$GaTe$_{2}$, as depicted in Fig.~\ref{fig1}(b), agree well with experiment \cite{Wu2024}. Additionally, from the DOS in Figs.~\ref{fig3}(b) and \ref{fig3}(d), sharp peaks begin to form near the Fermi level in both spin channels as the temperature decreases to around 300 K. These renormalized quasi-particle flat bands, as shown in Figs.~\ref{fig3}(a) and \ref{fig3}(c), result from the collective excitation of interacting electrons \cite{Yin2011,Xu2024,Kim2022}. Furthermore, the quasi-particle flat bands become sharper as the temperature decreases from 300 K to 50 K, indicating the presence of well-defined quasi-particles at low temperatures. Moreover, the electronic structures calculated in the ferromagnetic state show good agreement with recent ARPES measurements \cite{Lee2023,Wu2024}. In the spin-up channel of our calculated ferromagnetic state, hole-like bands appear near the $\Gamma$ point. As shown in Fig.~\ref{fig3}(a), these hole bands at 50 K closely correspond to the $\alpha$ and $\beta$ bands,represented by colored dashed lines extracted from APRES measurements \cite{Wu2024}. In the spin-down channel, the lambda-shaped dispersive bands along the $\Gamma$-$K$ path also consistent with the $\xi$ and $\zeta$ bands observed in APRES measurements, labeled by cyan and purple dashed lines in Fig.~\ref{fig3}(c). In addition, the spectral function in the spin-down channel reveals the presence of the $\gamma$ band along the $M$-$K$ path, which is in good agreement with ARPES results. Another APRES study has reported a low-dispersive band extending around -0.4 eV along the $\Gamma$-$M$ path, labeled as $\varepsilon$ in Fig.~\ref{fig3}(c) \cite{Lee2023}. This low-dispersive band is also clearly visible in our calculated spectral function of the spin-down channel. Compared to previous calculations for Fe$_{3}$GeTe$_{2}$ \cite{Xu2024}, the flat bands in Fe$_{3}$GaTe$_{2}$ are much closer to the Fermi level and become flatter.

Here, we focus on the discrepancy between DFT(+U) and DFT+DMFT calculations regarding the evolution of electronic structures in the presence of magnetism. In our work, the DFT+DMFT approach provides a more accurate description of magnetism compared to the other methods, primarily due to its consideration of dynamical spin fluctuations. In the DMFT framework, although long-range magnetic order does not emerge at high temperatures, local moments persist and fluctuate over time. In contrast, non-spin-polarized DFT calculations do not incorporate the concept of local moments. The differences in these theoretical frameworks can be quantitatively analyzed by comparing the magnetic moment and local magnetic moment. As shown in Table~\ref{tab1}, the magnetic moments of Fe atoms in the ferromagnetic state are larger in DFT+U calculations compared to standard DFT. However, in DFT+DMFT calculations at 50 K, the magnetic moments of Fe atoms, particularly Fe$2$, are significantly reduced. The average magnetic moment obtained from DFT+DMFT method at 50 K is 1.95 $\mu_{\rm B}$/Fe, which is slightly higher than the experimental saturation magnetic moment of 1.68 $\mu_{\rm B}$/Fe \cite{Zhang2022}. Moreover, the local magnetic moment obtained from DFT+DMFT is slightly larger than the ferromagnetic moment, highlighting the intrinsic spin fluctuation effects captured within the DMFT framework. Notably, our DFT+DMFT calculations indicate that the local moment remains nearly unchanged across the magnetic transition.

\begin{figure}
\begin{center}
\includegraphics[width=0.49\textwidth]{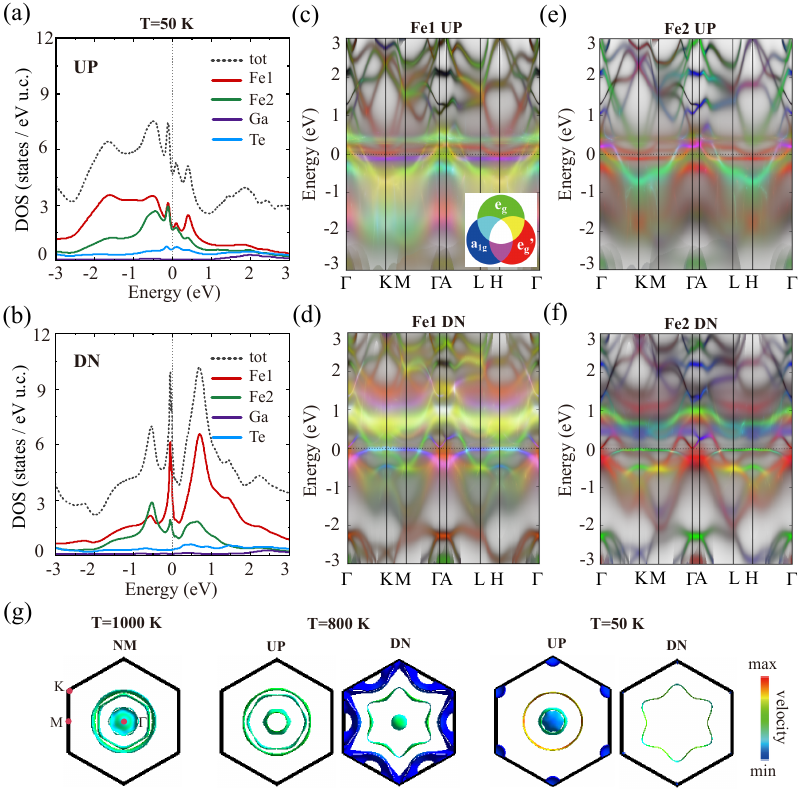}
\caption{The partial density of states at 50 K in the (a) spin-up channel and in the (b) spin-down channel. Orbital-resolved spectral function of Fe1 atoms at 50 K in the (c) spin-up channel and in the (d) spin-down channel. Orbital-resolved spectral function of Fe2 atoms at 50 K in the (e) spin-up channel and in the (f) spin-down channel. (g) Fermi surfaces and their velocity at different temperatures.}
\label{fig4}
\end{center}
\end{figure}

To gain a deeper physical understanding of these flat bands, we first investigated the electronic structures at the lowest calculated temperature of 50 K. The partial DOS, as shown in Figs.~\ref{fig4}(a) and \ref{fig4}(b), indicates that the flat bands primarily originate from the Fe-$d$ electrons. As depicted in Figs.~\ref{fig4}(c) to \ref{fig4}(f), in the spin-up channel, the flat bands are located at +0.1 eV and -0.15 eV near the Fermi level. In the spin-down channel, the flat bands are positioned at -0.01 eV, displaying a peak in the DOS. These renormalized flat bands can be considered heavy quasi-particles that emerge from collective excitations, similar to those found in Hund metals \cite{Yin2011,Georges2013,Georges2024}. Additional sharp peaks are observed at -0.5 eV in the spin-up channel and -0.06 eV in the spin-down channel, with contributions from Fe1, Fe2, and Te atoms at almost the same energy positions. These coinciding peak positions suggest hybridization between the localized Fe-$d$ orbitals and other itinerant electrons, akin to heavy fermion systems \cite{Stewart1984,Yang2008,Shim2007}. The spectral functions in Fig.~\ref{fig4} illustrate the hybridization characteristics of these flat bands, with the hybridization gap being most pronounced along the $\Gamma$-$A$ path. At last, as shown in Fig.~\ref{fig4}(g), the Fermi surface velocity decreases at low temperatures. Consequently, we deduce that Fe$_{3}$GaTe$_{2}$ exhibits heavy fermion behavior at low temperatures, similar to its reported sister material, Fe$_{3}$GeTe$_{2}$ \cite{Zhang2018,Zhao2021}.

We now address the potential mechanism connecting flat bands and magnetism. Previous experimental and theoretical studies suggest that the magnetic transition temperature can be enhanced through the manipulation of flat bands. In an ideal flat-band system, theoretical studies have rigorously demonstrated that even a small on-site Coulomb interaction can stabilize the ferromagnetic ground state \cite{Mielke1991, Tasaki1998}. Specifically, the Lieb lattice has been extensively studied as a model system for realizing ferromagnetism due to its distinctive flat bands. More recently, G. Bouzerar has shown that flat bands can promote room-temperature ferromagnetism in two-dimensional materials through dominant interactions between flat bands, using an $s$-$d$ Lieb model \cite{Bouzerar2023}. In recent decades, several flat-band materials have been experimentally identified, providing evidence that flat bands can influence ferromagnetic transitions. For example, the Kagome lattice material Fe$_{3}$Sn$_{2}$ exhibits highly degenerate electronic states at the Fermi level due to flat bands, making the system highly susceptible to ferromagnetic ordering even with a small on-site Coulomb interaction \cite{Lin2018}. Furthermore, the monolayer of graphitic C$_{4}$N$_{3}$ has been realized, which features the half-filled isolated flat bands and displays a ferromagnetic transition temperature of 241 K \cite{He2025}. Additionally, ferromagnetic transitions have been observed in other materials such as the carbon monolayer cyclic graphdiyne \cite{You2019}, the pyrochlore oxide Pb$_{2}$Sb$_{2}$O$_{7}$ \cite{Hase2023}, and the Lieb-like $sp^{2}$ carbon-conjugated covalent-organic framework \cite{Jiang2019}, where doping into the flat bands induces the ferromagnetic transition. Given the high ferromagnetic transition temperature in Fe$_{3}$GaTe$_{2}$, it is natural to consider its potential relationship with flat bands.

\begin{figure}
\begin{center}
\includegraphics[width=0.49\textwidth]{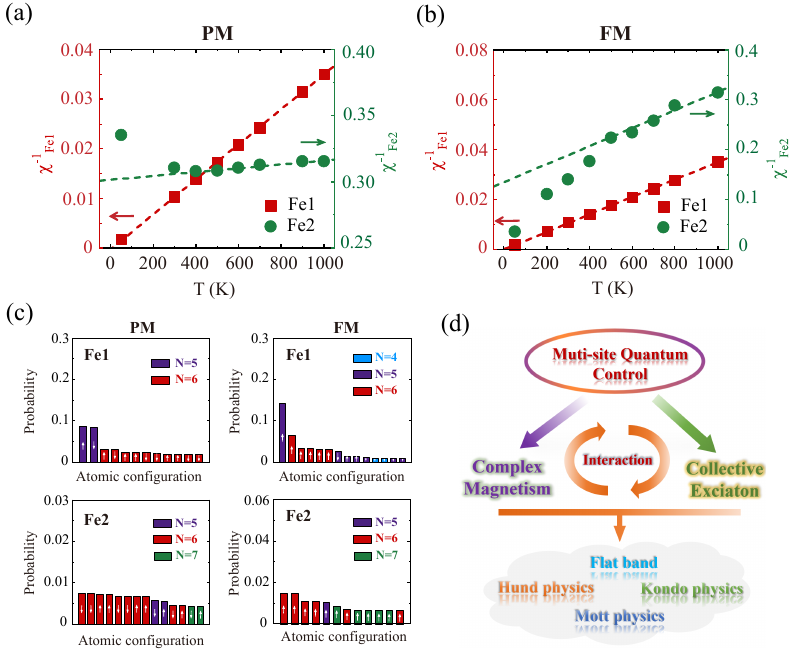}
\caption{The discrepancy in the behavior of Fe1 and Fe2 in Fe$_{3}$GaTe$_{2}$. (a) The inverse spin susceptibility versus temperature in PM calculations. (b) The inverse spin susceptibility versus temperature in FM calculations. (c) The most probable Fe atomic states obtained from the impurity solver in the high-temperature PM phase and low-temperature FM phase. In CTQMC simulations, the iron atom explores multiple atomic states over time within a given period. The notation `N' represents the occupation number of a given atomic configuration obtained from the impurity solver. The atomic states are ranked according to their probabilities, and the highest-probability Fe atomic states identified in our calculations are shown here. The $\uparrow$ arrow means $S_{z}$ is positive and the $\downarrow$ arrow means $S_{z}$ is negative. (d) A schematic diagram illustrating the interaction of flat bands and magnetism in multi-site correlated itinerant magnets.}
\label{fig5}
\end{center}
\end{figure}

To investigate this potential relationship, we calculated the spin susceptibility at varying temperatures. First, the spin susceptibility obtained from PM calculations, as shown in Fig.~\ref{fig5}(a), reveals that both Fe1 and Fe2 sites exhibit Curie-Weiss behavior over a broad temperature range. However, it is noteworthy that the spin susceptibility of Fe2 site deviates from the Curie-Weiss law below 400 K. This deviation may arise from the Fe-$d$ electrons forming quasi-particles with screening effects, such as Kondo screening \cite{Xu2022}. In contrast, within our calculated temperature range, the susceptibility of the Fe1 site in PM calculations continues to follow the Curie-Weiss law. Comparing the susceptibility behavior of Fe1 and Fe2 sites, we speculate that spin fluctuations are more pronounced at Fe2 site. Furthermore, magnetic DFT+DMFT calculations, as depicted in Fig.~\ref{fig5}(b), reveal that the Curie-Weiss law holds above the Curie temperature. Below approximately 400 K, a notable increase in the spin susceptibility of the Fe2 site is observed, accompanied by a deviation from the Curie-Weiss law. Interestingly, this deviation follows an opposite trend compared to that observed in paramagnetic DFT+DMFT calculations. Such behavior is expected, as spin susceptibility generally increases as the system approaches the ferromagnetic transition in real materials. However, a surprising observation is that the susceptibility of the Fe1 site continues to follow the Curie-Weiss law across the entire investigated temperature range. It is well known that near a ferromagnetic phase transition, short-range magnetic order and significant quantum fluctuations render the mean-field-based Curie-Weiss model inadequate \cite{Kubler2009}. These effects are not fully captured within single-site DFT+DMFT calculations, which likely explains the observed Curie-Weiss-like susceptibility of the Fe1 site. This interpretation is further supported by the fact that the larger magnetic moment and higher susceptibility of the Fe1 site help suppress spin fluctuations of Fe1 atoms. Conversely, the lower susceptibility and stronger spin fluctuations at the Fe2 site contribute to the observed deviation in its susceptibility behavior in magnetic DFT+DMFT calculations.

Furthermore, our results provide insights into the influence of flat bands on magnetic anisotropy. In previous studies on the isostructural sister compound Fe$_{3}$GeTe$_{2}$, PMA has been shown to be closely associated with flat bands along the $\Gamma$–$A$ and $K$–$H$ directions, where a positive contribution to PMA is observed around the $K$ point, while a negative contribution is found around the $\Gamma$ point. Even slight tuning of these flat bands relative to the Fermi level can significantly impact magnetic anisotropy \cite{Park2020}. In our spectral function calculations, flat bands are present near the $\Gamma$ and $K$ points. Referring to previous studies incorporating spin-orbit coupling effects in Fe$_{3}$GaTe$_{2}$ \cite{Lee2023}, we recognize that these flat bands may undergo splitting due to spin-orbit coupling, contributing to the PMA in this material. Furthermore, magnetic anisotropy in Fe$_{3}$GaTe$_{2}$ is strongly influenced by the magnetic interactions between Fe1 atoms along the $c$-axis \cite{Ruiz2024, Lee2023}. Our orbital-resolved spectral function calculations, as shown in Fig.~\ref{fig4}(d), reveal a significant presence of the Fe1-$a_{1g}$ ($d_{z^2}$) orbital character within the flat bands at the Fermi level, particularly near the $K$ point. This orbital character at $K$-point is known to contribute to the positive PMA in Fe$_{3}$GeTe$_{2}$ \cite{Park2020}, suggesting a similar mechanism in Fe$_{3}$GaTe$_{2}$.

The discrepancy between the behavior of Fe1 and Fe2 arises from their distinct local crystal environments \cite{Xu2024}. As shown in Fig.~\ref{fig5}(c), the highest atomic probabilities from the DMFT impurity solver indicate that Fe1 and Fe2 sites exhibit significant local moments and considerable spin fluctuations in the high-temperature PM phase. At low temperatures, the highest probabilities for Fe1 and Fe2 sites are predominantly positive values of $S_{z}$, indicating a spontaneous symmetry breaking to ferromagnetic order. Additionally, Fe1 site tends to be concentrated in a single charge state, characteristic of Mott-type behavior \cite{Georges2024}. In contrast, the histogram for Fe2 site spans multiple charge states, indicative of Hundness \cite{Yin2011,Georges2024,Deng2019}. For simplicity, Fig.~\ref{fig5}(d) proposes a new efficient avenue for developing applicable spintronic devices. The proposed multi-site systems provide more opportunities to control novel quantum states. A critical issue is the production of flat bands, whether through Hund, Mott, or Kondo mechanisms \cite{Checkelsky2024}. Among these, Hund's coupling is recognize to be important, and fortunately, it is intrinsic and widely present in itinerant correlated metals. The flat bands can enhance the DOS at the Fermi level, raising the possibility of forming the ferromagnetic order. Once the flat bands emerge and drive the magnetic phase transition, the resulting long-range magnetic order rapidly reduces spin fluctuations and promotes flat band formation through positive feedback. We recognize that more high-$T_{\rm C}$ candidate spintronic devices could be discovered through this microscopic mechanism.

\section{CONCLUSION}

In conclusion, we have systematically investigated the electronic and magnetic properties of Fe$_{3}$GaTe$_{2}$ using DFT+DMFT calculations. Our results reveal a magnetic transition above room temperature, with the calculated spin susceptibility exhibiting Curie-Weiss behavior at high temperatures. We observe spectral weight transfer between spin channels, underscoring the significance of many-body interactions in this system. Notably, our study demonstrates the emergence of quasiparticle flat bands at low temperatures, stabilized by long-range magnetic order and exhibiting heavy fermion characteristics. The presence of flat bands near the Fermi level offers a promising avenue for optimizing the Curie temperature and enhancing magnetic anisotropy. These findings provide crucial insights into the correlated magnetism of Fe$_{3}$GaTe$_{2}$ and highlight a promising approach for discovering more practical spintronic devices through the precise tuning of flat bands.

\begin{acknowledgments}
This work is supported by the National Natural Science Foundation of China (Grant Nos. 12204033, 52371174, U2230401), the Fundamental Research Funds for the Central Universities (Grant No. FRF-TP-22-097A1), the State Key Lab of Advanced Metals and Materials (Grant No. 2022Z-13) and the Young Elite Scientist Sponsorship Program by BAST (Grant No. BYESS2023301). Numerical computations were performed on Hefei advanced computing center.
\end{acknowledgments}

\end{document}